\documentclass{epl}

\title{The growth of a Super Stable Heap :
\\an experimental and numerical study}
\author{N. Taberlet \and P. Richard \and E. Henry \and R. Delannay}
\institute{
Groupe Mat\'eriaux et Mati\`ere Condens\'ee, Universit\'e Rennes 1
\\   Bat 11A Campus Beaulieu, 35042 Rennes Cedex, France\\
}
\pacs{45.70.Ht}{Avalanches}
\pacs{45.70.Mg}{Granular flow}
\pacs{83.10.Mj}{Molecular dynamics in rheology}

\begin{document}

\maketitle

\begin{abstract}
We report experimental and numerical results on the growth of a super stable heap (SSH). Such a regime appears for flows in a thin channel and for high flow rate : the flow occurs atop a nearly static heap whose angle is stabilized by the flowing layer at its top and the  side wall friction.
The growth of the static heap is investigated in this paper.
A theoretical analysis inspired by the BRCE formalism predicts the evolution of the growth process, which is confirmed by both experiments and numerical simulations.
The model allows us to link the characteristic time of the growth to the exchange rate between the "moving" and "static" grains.
We show that this rate is proportional to the height of the flowing layer even for thick flows. The study of upstream traveling waves sheds new light on the BCRE model.
\end{abstract}

\section{Introduction}

The dynamics of granular media appears to be very different from that of usual fluids~\cite{Jaeger1996}. In particular such media have the ability to remain static (i.e. a solid-like behavior) at a non-zero tilt angle. When this tilt angle exceeds a critical value a surface flow is observed : most of the pile remains static except for a relatively thin rolling layer~\cite{Komatsu01, Aradian99}. A granular medium can thus display both solid-like and fluid-like behaviors. Very recently~\cite{PRL}, it has been shown that when the granular medium is confined between two vertical glass walls, the angle of the free surface increases with increasing input flow rate. Unusually steep granular heaps (Super Stable Heap - SSH) can then be obtained. This letter presents an experimental and numerical study on the growth of such heaps which allows one to characterize the exchange rate between the fluid-like  and the solid-like phases.

\section{Summary of the previous work on Super Stable Heaps}

\begin{figure}[htbp]
\begin{center}
\resizebox{14.4cm}{!}{\includegraphics*{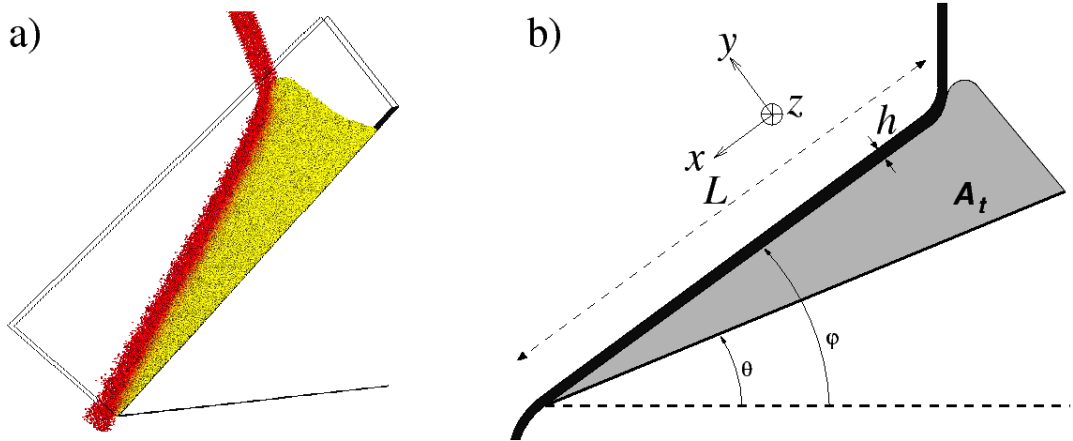}}
\caption{a) (In color online) Snapshot of the flow from numerical simulation. Velocity color-scale b) Sketch of the transient regime. The dark part corresponds to moving grains, the gray part to static grains.}
\label{fig_schema}
\end{center}
\end{figure}

All experiments presented in this paper were performed in the same channel as in our previous work~\cite{PRL}. Our experimental set-up consists of a thin three-dimensional open channel that could be inclined from horizontal to an angle $\theta$ ({figure}~{\ref{fig_schema}a} shows a similar system used in our simulation). Although $\theta$ could range from $0^\circ$ to $90^\circ$, all experiments were conducted in a steep configuration in which $\theta$ was always set to be higher than the angle at which all the grains flow out of the channel after the feeding was cut off. Experiments were performed using Fontainbleau sand (round shaped, 0.3-0.5 $\mu$m). Granular material was continuously poured on a rough bottom placed between two transparent vertical glass plates through a hopper whose aperture precisely controlled the input flow rate $Q_{in}$ (mass per unit of time and per unit of width).
The rough bottom was made of grains glued to a flat rigid substrate. The gap between the two vertical plates $W$ was varied from 5 mm up to 20 mm and the length of the channel, $L$, could by varied by moving the hopper. For further details see~\cite{PRL}.

For low flow rates we observed classical flows on the rough bottom of the channel but a new regime, occuring for high flow rates, was investigated. In this regime, a static wedge-like heap gradually forms: some of the flowing grains get trapped and contribute to the growth of the static heap. Let us mention here that the grains are not completely static but can experience a slow creeping motion. For simplicity, these grains will however be refered to as static. The system eventually reaches a steady state consisting of a uniform flow atop a static heap. Let us mention here that although the flow is uniform when averaged over time, there can exists local fluctuations of the flow height, $h$, sometimes corresponding to upstream traffic waves described in~\cite{Louge01}. Following the definition in~\cite{PRL}, $h$ was measured using the velocity profiles. When the feeding is cut off, the heap slowly erodes and falls apart. This shows that the heap is dynamically stabilized by the flow at its top. One remarkable feature exhibited by such heaps is their unusual steepness (up to $60^{\circ}$ in our experiments) and this is why they are referred to as Super Stable Heaps. {Figure}~{\ref{fig_schema}b} is a sketch of the regime described here. The SSH appears in light gray and the flowing layer in black.
For given values of $Q_{in}$ and $W$, the system "chooses" an angle $\varphi$ for the free surface and the corresponding height of the flow $h$. The values of $\varphi$ and $h$ depend only on the input flow rate, $Q_{in}$, and on the channel width, $W$, and not on the tilt angle $\theta$ nor on the channel length, $L$. A simple depth-averaged model provided us with a simple law linking the angle of the SSH in the final steady state, the height of the flow and the channel width. Writing the momentum balance leads to:
\begin{equation}
\tan \,\varphi_{\infty} \, =\, \mu_i \,+\, \mu_w\, \displaystyle{\frac{h}{W}}
\label{equ_PRL}
\end{equation}
where $\mu_i$ and $\mu_w$ are two positive material constants.

Our previous study focused on the steady state and identified the mechanisms responsible for the existence of SSH. The present paper focuses on the growth of the SSHs i.e. the time transient leading to the steady state. The study of this transient regime is of great interest on its own and can moreover provide one with quantitative information on the exchanges existing between static and moving grains when granular material is flowing on a static heap. Our experiment might then shed some light on the BCRE model~\cite{BCRE}.

\section{Preliminary Observations}

\begin{figure}[htbp]
\begin{center}
\resizebox{7.2cm}{!}{\includegraphics*{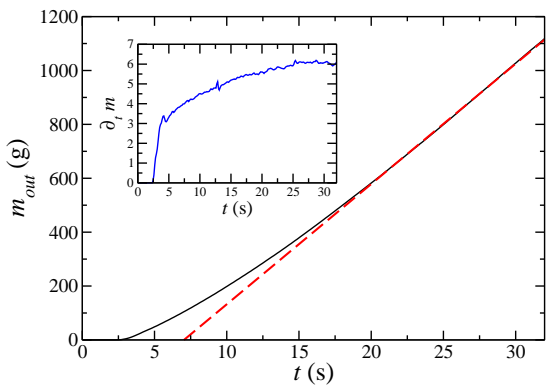}}
\caption{Plot of the output mass $m_{out}$ vs. time measured experimentally. Inset: $Q_{out}=\partial_t m$ vs. time}
\label{fig_Qout}
\end{center}
\end{figure}

Before discussing the results let us mention two simple observations that simplified the detailed study of the growth of the heap. First, we observed that the growth was a "slow" process: 
among the many grains entering the channel, only very few of them actually contributed to building the heap. {Figure}~{\ref{fig_Qout}} is a plot of $m_{out}$, the mass of grains having exited the channel, vs. time and shows that the output rate $Q_{out}$ slowly tends toward a constant. This is clear on the inset which is a plot of $Q_{out} = \partial_t m_{out}$. One can see that even after 30 seconds, a period much greater than the time it takes a grain to flow through the channel (typically 1s), the steady regime is not yet reached. Let us mention that the variation of the height, $h$, was found to be very small during the growing process.

The second preliminary observation is that the surface of the heap remains flat throughout the whole duration of the growth. That is to say that the heap maintains a linear profile while its slope changes over time. One can therefore define a single angle for the whole heap, $\varphi_t$, at any given time $t$ (see {figure}~{\ref{fig_schema}b}).

The free surface presents no curvature in the $z$ direction, meaning that a side-view is sufficient to describe the system.
For technical reasons the surface area of the heap, $A_t$, was easier to measure than the angle $\varphi_t$.
In both our experimental system and our simulation, the channel was long enough to allow for the neglect of the surface area of the flowing layer relative to that of the static heap (less than 5\%). The surface area of the whole heap (SSH and flowing layer) was then assumed to be equal to the surface area of the SSH. This allowed for easier measurements of $A_t$. Given these observations, a simplified geometrical expression for the surface area of the heap can be given, as long as $\varphi_t - \theta$ is not too large (i.e. < 30°):
\begin{equation}
A_t \, =\, \frac{L^2}{2} \, (\varphi_t - \theta)
\label{equ_geometric}
\end{equation}

\section{Theory}

As mentioned above, the flow consists of a layer moving rapidly atop a static heap.
Since a flowing grain can be trapped and conversely a static grain can start moving after a collision, there exist exchanges between the two species of grains.
The distinction between two species of grains is the basis of the BCRE model \cite{BCRE}. In this model, the height profile of the flowing layer obeys a convective diffusion equation, and the exchanges between the two species are ruled by a local law: accretion is proportional to the difference between the neutral angle (i.e. the angle at which erosion compensates accretion) and the local slope of the flow. The positive proportionality constant, denoted $\Gamma$ is called the exchange rate and is proportional to the flow height, $h$~\cite{BCRE}.
However, it was suggested in the literature~\cite{Boutreux98} that this should not hold for thick flows ($h \, >$ 5-10 $d$), for which $\Gamma$ should be proportional to $v_{up}$, the velocity of upstream traveling waves, the existence of which has been reported in the literature~\cite{Louge01}.


Our idea was to use a law similar to that of the BCRE model in order to describe the overall growth rate of the SSH. In our experiments, the neutral angle was none other than $\varphi_{\infty} $, the angle of the SSH in the final steady state since then the angle of the heap remains constant, meaning that erosion exactly compensates accretion. Therefore, we assumed that the surface area of the SSH follows the law:

\begin{equation}
\partial_t A_t \,=\, \Gamma (\varphi_{\infty}-\varphi_t)  \,\,\, ,  \, (\Gamma >0)
\label{equ_mon_BCRE}
\end{equation}

\noindent Using {equation}~{(\ref{equ_geometric})} and {(\ref{equ_mon_BCRE})} leads to:

\begin{equation}
A_t \,=\, A_{\infty} \, (1- e^{-\frac{t}{\tau}}) ,   \mbox{where}\,\, \tau \,=\, \frac{L^2}{2 \Gamma}
\label{equ_At}
\end{equation}

\noindent This model predicts an "1-exp" growth and provides us with a characteristic growth time, which appears to be inversely proportional to the exchange rate, $\Gamma $.

\section{Simulation methods}

Numerical simulations were performed using a Molecular Dynamics (MD) method, a.k.a. Discrete Element Method (DEM). This type of simulation has been widely used in the past and has proven to be a very reliable method~\cite{Silbert01}. 
{Figure}~{\ref{fig_schema}a} presents a snapshot of the flow obtained with our simulation.
In the simulation, we used a dashpot-spring force model : $F^n_{ij} \,=\, k_n \,\delta_{ij} + \gamma_n \,\partial_t\delta_{ij}$ for the normal force and a Coulomb friction law for the tangential force : $F^t_{ij} = min(\mu F^n_{ij} , \gamma_t V^s_{ij})$, where $\delta_{ij}$ is the overlap between two colliding particles, $k_n$ a spring constant, $\gamma_n$ a viscous damping constant, $\mu$ a friction coefficient, $V^s_{ij}$ the sliding velocity of the contact, $\gamma_t$ a visous regularization constant. The following values were used : particle radius $R$=4 mm, mass=0.16 mg, $k_n=40000\mbox{ N.m}^{-1}$, $\gamma_n=1.2\mbox{ s}^{-1}$ and $\gamma_t=5\mbox{ s}^{-1}$, leading to a coefficient of restitution of 0.4. Different values of the latter were tried but did not affect the flow properties, which was expected since in dense assemblies of grains the effective coefficient of restitution is close to zero, regardless of the material properties.


Because of the nature of the phenomenon, it was not possible to use periodic boundary conditions.
The channel is constantly fed by releasing grains from a virtual hopper.
The rough bottom of the channel consisted of immobile particles. The collisions against the side walls were treated like particle-particle collisions (identical simulation parameters $\mu$, $k_n$, $\gamma_n$ etc) with one of the particle having infinite mass and radius, which mimics a large flat surface. The granular material was made slightly polydisperse ($\pm20\%$ in size) in order to avoid crystallization. Our simulations typically contained a very large number of particles, between $10^5$ and $10^6$, depending on the size of the channel and on the input flow rate $Q_{in}$ and were run for typically $10^7$ time steps.

\section{Experimental and Numerical Results}

The time evolution of the surface area, $A_t$, was studied both experimentally and numerically. In the experiments, $A_t$ was measured using imaging software. As for the simulations, $A_t$ was computed from the number of grains belonging to the SSH.

\begin{figure}[htbp]
\begin{center}
\resizebox{14.4cm}{!}{\includegraphics*{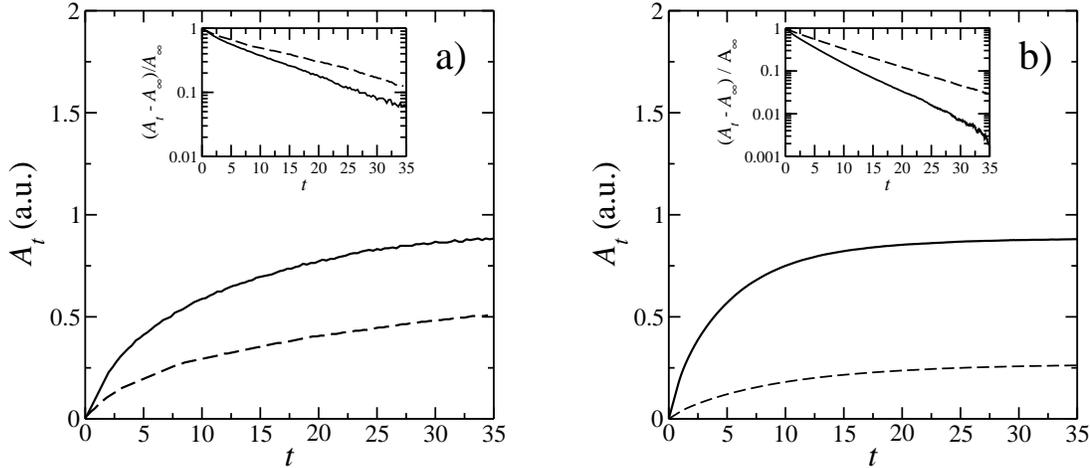}}
\caption{Surface area $A_t$ vs. time. Insets: semi-log plot of $ (A_{\infty}-A_t)/A_{\infty}$. a): experiments. solid line: $L=1\,m,  \, W=5  \, mm, \, Q=2.4  \, g.s^{-1}.mm^{-1}$, dashed: $L=1\,m, \, W=10 \, mm, \, Q=11.1 \, g.s^{-1}.mm^{-1}$.  b) numerics. solid line: $L=200\,R, \, W=10\,R, \, Q=225 \, part.s^{-1}.mm^{-1}$, dashed: $L=200\,R, \, W=20\,R, \, Q=650  \, part.s^{-1}.mm^{-1}$}
\label{fig_At}
\end{center}
\end{figure}

\noindent {Figure}~{\ref{fig_At}} shows plots of the surface area of the SSH vs. time measured experimentally and numerically for various values of the control parameters $L$, $W$ and $Q_{in}$. The evolution seems to follow an 1-exp law, which is confirmed by the right-hand side of the figure. The insets on {Figure}~{\ref{fig_At}} show semi-log plots of $(A_{\infty} - A_t)/A_{\infty}$ vs. $t$ and presents excellent agreement with {equation}~{(\ref{equ_At})} since all data follow a linear law. Indeed, both experimental and numerical data, for a wide variety of parameters, present a remarkable agreement with the "1-exp" law. One can therefore measure $\tau$ and compute the value of the exchange rate, $\Gamma $, for various values of the control parameters.

\begin{figure}[htbp]
\begin{center}
\resizebox{14.4cm}{!}{\includegraphics*{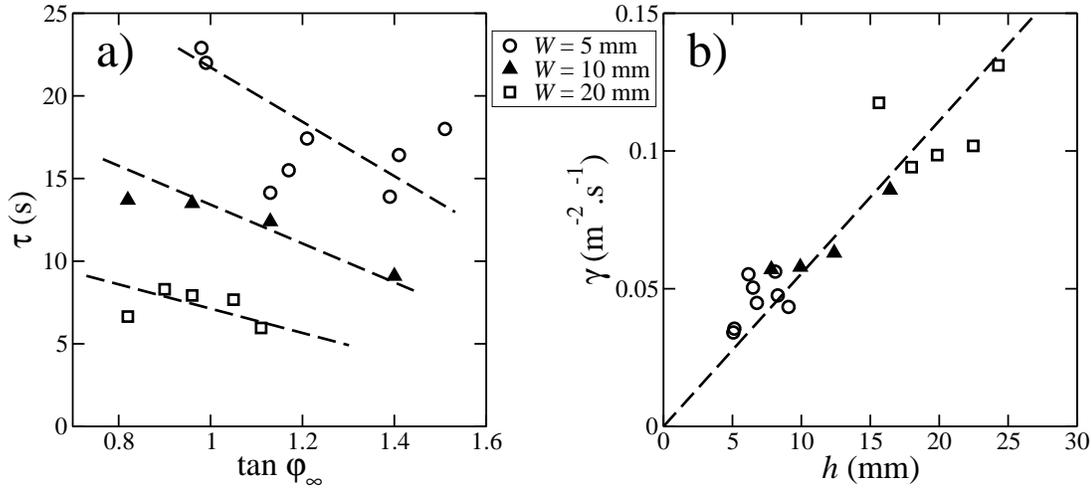}}
\caption{a) Characteristic growth time $\tau$ vs. tan $\varphi_{\infty}$ measured experimentally. b) exchange rate $\gamma$ vs. flow height, $h$}
\label{fig_tau}
\end{center}
\end{figure}

In this last section, we would like to comment on the exchange rate $\Gamma$. As mentioned above the value of $\Gamma$ can be extracted from the characteristic time, $\tau$, using: $\tau = L^2/ (2 \, \Gamma) $. The $L^2$ scaling was expected since the properties that were monitored in this study are surface areaand checked numerically. A more relevant, geometry-independent exchange rate, $\gamma$, can then be defined: $\gamma \,=\, 2 \Gamma / L^2 = 1/ \tau$. Measurements of $\tau$ were performed for various values of $W$ and $Q_{in}$, allowing for the study of $\tau$ as a function of the angle $\varphi_{\infty}$. {Figure}~{\ref{fig_tau}a} is a plot of $\tau$ vs tan $\varphi_{\infty}$. The straight lines do not come from a fit but only help to visualize the general trend.

The first conclusion that can be drawn from this plot is that $\tau$ depends on the channel width $W$. That is to say that for a given angle $\varphi_{\infty}$, $\tau$ varies when $W$ is changed. Namely, the thinner the channel, the larger the characteristic time.
In our previous work on SSHs we showed that there exists a simple relationship between $\varphi_{\infty}$, $h$ and $W$ ({equation}~{(\ref{equ_PRL})}). In the light of this previous result, one can interpret the data in a new fashion.
For a given $\varphi_{\infty}$, $\tau$ decreases with increasing $W$, as for a given $W$, $\tau$ decreases with increasing $\varphi_{\infty}$, meaning in both cases that $\gamma$ increases with increasing $h$.
The function $\tau(\varphi_{\infty},W)$ can be expressed as the new function $\gamma(h)$, which is easier to interpret. 
As mentioned above, the heigth of the flow remains constant during the growth, and can therefore be computed as follows.
The values of the coefficients $\mu_i$ and $\mu_w$ having been determined in our previous work ($\mu_i$=tan $23.3^\circ$ and $\mu_w$=tan $33.7^\circ$), one can then compute the flow height in the steady regime of each run using {equation}~{(\ref{equ_PRL})}. 
One can then plot $\gamma$ as a function of $h$. {Figure}~{\ref{fig_tau}b} presents such a plot and shows a decent data collapse. $\gamma$ is found to be a linear function of the flow height $h$.
In agreement with theoretical work based on the BCRE model~\cite{Aradian99}, the exchange rate is proportional to the height of the flow but in our case, no saturation was observed.

\begin{figure}[htbp]
\begin{center}
\resizebox{14.4cm}{!}{\includegraphics*{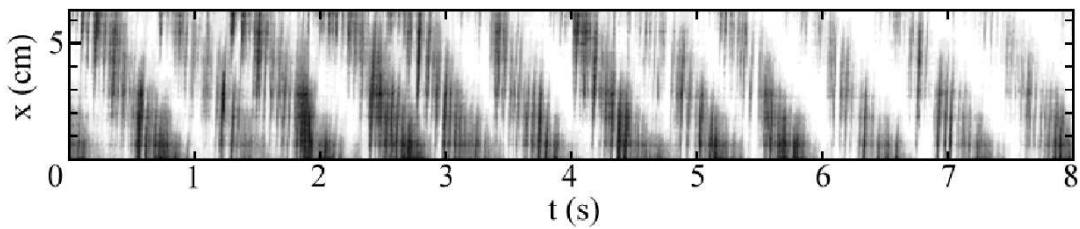}}
\caption{Space-time diagram of the fluctuations of the height of the flowing layer, $h(x,t)-h$.}
\label{fig_XT}
\end{center}
\end{figure}

Although $\gamma$ is clearly a linear function of $h$, it remains interesting to check whether $\gamma$ can be expressed as a function of $v_{up}$, the velocity of upstream traveling waves. In order to compute this velocity, we measured the fluctuations of the height, $h(x,t)$, in the final steady state.
{Figure}~{\ref{fig_XT}} is a grey-scale space-time diagram of $h(x,t)-h$. On this figure, a dark spot indicates a local height, $h(x,t)$, greater than the average i.e. a hump, whereas a white spot corresponds to a depression.

\noindent Two patterns can be observed on {figure}~{\ref{fig_XT}} : on one hand, quasi-vertical narrow lines, on the other hand, wide oblique stripes. The former correspond to perturbations "floating" at the free surface (like a object floating on a river) and therefore moving downward at the velocity of the free surface, $v_{surface}$, whereas the latter corresponds to upward traveling waves. The main direction of both kinds of stripes (i.e. $v_{surface}$ and $v_{up}$) can be extracted using an angular FFT method or directly measured by fitting them with straight lines. The narrow lines (corresponding to downward motion) appear to be less oblique than the wide stripes (corresponding to upward waves), meaning that $v_{surface} > v_{up}$ (on {figure}~{\ref{fig_XT}} : $v_{surface}$= 82 cm/s and $v_{up}$ =11.8 cm/s). In other words, the upward travelling waves travel slowly compared to the characteristic velocity of the flow. Secondly, the spacing between the wide stripes seems to be regular, indicating that there exists a well-defined wave-length. The origin of this wave-length is unclear and might be related to the phenomena responsible for the onset of the upward travelling wave.
The velocity $v_{up}$ was measured for different values of the control parameters but no simple relationship could be found between $\tau$ and $v_{up}$. 
This last result seems to indicate that even for the thick flows we observed ($h\simeq 50 d$), the exchange rate is not proportional to $v_{up}$.
Let us mention here that the use of $v_{up}$ in the BCRE model has already been questioned in the literature~\cite{Rajchenbach02}.

\section{Conclusion}

The growth of the Super Stable Heap was studied both experimentally and numerically (using the Molecular Dynamics method). Among the many grains flowing at the surface of the static heap, a few get trapped and contribute to the formation of the heap. During the growing process the free surface of the flow was found to remain flat. This allowed us to introduce a model for the accretion/erosion process, similar to the BCRE model~\cite{BCRE}. A global accretion/erosion law predicted the time evolution of the surface area of the heap that was confirmed by both experiments and simulations. The model provides one with a characteristic time which is a direct measurement of the exchange rate, $\gamma$.
We showed that $\gamma$ is proportional to the height of the flow, $h$, even for thick regimes ($h\simeq 50 d$). However, no simple relationship could be established between $\gamma$ and $v_{up}$, the velocity of upward traveling waves (traffic waves).
It was suggested in~\cite{Boutreux98} that they should be proportional for thick flows. This disagreement might indicate that $\gamma$ and $v_{up}$ are never proportional or that there exists a fundamental difference between flows on SSHs and on classic heaps. The relationship between $\gamma$ and $v_{up}$ should be investigated experimentally in the latter case.

\end{document}